\documentclass[
pra, 
aps,
twocolumn,
superscriptaddress,
longbibliography,
floatfix,
notitlepage]{revtex4-2}

\usepackage[utf8]{inputenc}
\usepackage[english]{babel}
\usepackage[T1]{fontenc}
\usepackage{lmodern}

\usepackage{amsmath,amsfonts,amssymb,amsthm,bm,times,dcolumn}

\usepackage{microtype}
\usepackage{braket}
\usepackage{gensymb} % for example \degree symbol can be used!
\usepackage{physics}

\usepackage[colorlinks={true}, citecolor={blue}, filecolor={blue}, 
linkcolor={blue}, urlcolor={blue}]{hyperref}
\usepackage{graphicx,color}
\usepackage[caption=false]{subfig}

\usepackage{soul}
\usepackage[normalem]{ulem}

\begin{document}
	
\preprint{APS/123-QED}

%\title{Work harvesting by q-deformed statistical mutations in quantum Otto cycle}
%\title{The effect of the q-deformation on the work efficiency in a bosonic quantum Otto cycle}	
%\title{q-deformed quantum Otto engine}
%\title{Creating quantum Otto cycle with q-deformation of the working substance}
\title{Powering quantum Otto engines only with q-deformation of the working substance}

\author{Fatih Ozaydin}
\email{fatih@tiu.ac.jp}
\affiliation{Institute for International Strategy, Tokyo International University, 1-13-1 Matoba-kita, Kawagoe, Saitama 350-1197, Japan}
\affiliation{CERN, CH-1211 Geneva 23, Switzerland}
%\affiliation{Department of Information Technologies, Isik University, Sile, Istanbul, 34980, Turkey}
	
\author{\"{O}zg\"{u}r E. M\"{u}stecapl{\i}o\u{g}lu}
%\email{omustecap@ku.edu.tr}
\affiliation{Department of Physics, Ko\c{c} University, 
Sar{\i}yer, \.Istanbul, 34450, Türkiye}
\affiliation{TÜB\.ITAK Research Institute for Fundamental Sciences, 41470 Gebze, Türkiye}

\author{Tuğrul Hakioğlu}
%\email{hakioglu@itu.edu.tr}
\affiliation{Energy Institute,  \.Istanbul Technical University, Sar{\i}yer,  \.Istanbul, 34467, Türkiye}
\affiliation{Department of Physics Engineering, \.Istanbul Technical University, Sar{\i}yer,  \.Istanbul, 34467, Türkiye  }
\affiliation{Department of Physics, Northeastern University, Boston, MA 02115, USA}
\date{\today}

\begin{abstract}
We consider a quantum Otto cycle with a $q$-deformed 
quantum oscillator working substance and classical thermal baths. 
We investigate the influence of the quantum statistical deformation
parameter $q$ on the work and efficiency of the cycle. In usual quantum Otto cycle, 
a Hamiltonian parameter is varied during the quantum adiabatic
stages while the quantum statistical character of the working substance remains fixed. 
We point out that even if the Hamiltonian parameters are not changing, 
work can be harvested by quantum statistical changes of the working substance. 
Work extraction from thermal resources using quantum statistical mutations of the
working substance
makes a quantum Otto cycle without any classical analog. 
\end{abstract}

\maketitle

%%%%%%%%%%%%%%%%%%%%%%%%%%%%%%%%%%%%%%%%%%%%%%%%%%%%%%%%%%%%%%%%%%%%%
\section{Introduction}\label{sec:introduction}
%%%%%%%%%%%%%%%%%%%%%%%%%%%%%%%%%%%%%%%%%%%%%%%%%%%%%%%%%%%%%%%%%%%%%
Quantum Heat Engines (QHEs) are devices that can harvest work using a quantum working 
substance between hot and cold reservoirs~\cite{scovil59,geusic67}.
After the foundations of the quantum engines have been established, many researchers 
have devoted intense theoretical and experimental efforts to find new breakthroughs~\cite{quan2007,altintas2014,altintas2015,ross2014,naseem2019,Hugel2002,
kosloff2014,barontini2019,abah2012,martinez2016,ross2016,tuncer2019work,dag2019temperature,Cakmak:22,PhysRevLett.123.240601}. Enhancement of work and efficiency of such quantum machines, together with 
exploring their fundamental bounds, are among the significant goals of the emerging 
field of quantum thermodynamics. For that aim, non-linear, many-body, fermionic or 
bosonic working systems have been studied to reveal their differences and relative 
advantages~\cite{quan2007}.
Here, we contribute to these research endeavors by addressing two questions. 
First, how the engine performance depends on quantum statistics in general 
if we mutate the particle statistics? 
Second, can we consider quantum statistics as another control parameter such that 
if all the system parameters remain the same, we can harvest work from a heat bath by 
only changing the quantum statistics of the working substance?

In the 1970s, the concept of deformed algebras was first initiated as a generalization of Weyl-Heisenberg algebra~\cite{arik1975operator,arik1976hilbert}.
The theory of the $q$-oscillators was stated previously by Mcfarlane, and Biedenharn ~\cite{biedenharn89,mcfarlane89}. 
Since then, $q$-deformation has been considered in various research areas
including 
statistical physics and quantum information~{\color{black}\cite{altintas2014constructing}}, 
nuclear and atomic physics~{\color{black}\cite{bonatsos1992generalized}}, 
thermodynamics~{\color{black}\cite{naseri2022}}, 
open quantum systems and optomechanical systems~{\color{black}~\cite{kundu2022}}.

It is pointed out that there is a correspondence between $q$-deformed Heisenberg 
algebra and effective non-linear interaction of the cavity mode~\cite{doi:10.1080/09500349214550981}, 
and an isomorphism between the $q$-deformed harmonic oscillator and an anharmonic oscillator model was discovered~\cite{PhysRevA.47.2555}. Physical realization of the deformation parameter $q$ has 
been searched for heavily, among which are the quantum Yang-Baxter equation~\cite{zhong1993}, 
deformed Jaynes-Cummings model~\cite{chai1990}, quantum phase problem~\cite{ellinas1992,
hakioglu1998,hakioglu1998_2}, relativistic $q$-oscillator~\cite{kasimov1991,arik1992}, 
Morse oscillator~\cite{cooper1995}, and Kepler problem~\cite{dayi1995}. 
Deformed algebras have been explored by subjecting the non-deformed ones to non-linear invertible transformations~\cite{argonne1991quantum,curt1990,haki1996}.
The $q$-deformation parameter was considered for deriving generalized uncertainty and information relations~\cite{van1984generalized}, Tsallis entropy, and other relative entropy measures~\cite{naudts2011generalised,borges1998family,LANDSBERG1998211}.

In atomic and nuclear physics, $q$-deformation was considered from theoretical and experimental perspectives~\cite{bonatsos1992generalized,bonatsos2002deformed,georgieva2011q,
Altintas_2012q,hammad2019q,jafarizadeh2018study,Boutabba2022}.
It was considered for obtaining generalizations of quantum spin chains with exact valence-bond ground states~\cite{doi:10.1142/S021797929400155X}, and
self-localized solitons of $q$-deformed quantum systems have been recently 
explored~\cite{bayindir2021self}. Deformed algebra is also used in 
open quantum systems to show the relationship between the efficiency of QHEs 
and the degree of the non-Markovianity cycle processes~\cite{naseri2022}. 
One of the recent works showed that two linearly coupled $q$-deformed cavities 
could be tuned to provide enhancement of non-classical phenomena~\cite{kundu2022}.

In one of the earliest works on $q$-deformed quantum information, entanglement and noise 
reduction techniques were studied between $q$-deformed harmonic oscillators~\cite{Dattoli1996}.
Nonclassical properties of noncommutative states~\cite{dey2015q,berrada2019noncommutative}, coherent and cat states in Fock representation~\cite{fakhri2021q,fakhri2020nonclassical} and 
entanglement in nonlinear quantum systems were analyzed in q-deformed settings~\cite{berrada2012bipartite}.
Quantum states and logic gates were defined for two- and three-level 
$q$-deformed systems~\cite{filippov1991harmonic,altintas2014constructing,altintas2020q}, 
and $q$-deformed relative entropies were studied in quantum metrology~\cite{hasegawa2006quantum}.

Our work presents a unique perspective on the quantumness of heat engines, 
which reflects the genuine quantum statistical character of the working system in 
 harvesting work from classical thermal resources. The usual method to characterize the 
quantum nature of a heat engine is to look for quantum-enhanced performance over its classical 
analog. Our case is another yet more direct reflection of the quantumness of a heat engine 
as the cycle mechanism, which is based upon harvesting work by changing 
the quantum statistical character of its working substance, has no classical analog. 

This paper is organized as follows: Sec.~\ref{sec:model} introduces the $q$-deformed 
quantum oscillator as the working system of the quantum Otto cycle. 
Sec.~\ref{sec:quantumOttoCycle} presents the necessary tools to construct a $q$-deformed heat 
engine by discussing the fundamental thermodynamic quantities such as entropy and internal 
energy. Then, equipped with the theoretical tools presented in previous sections, we present our results in Sec.~\ref{sec:results}, where 
we explicitly show how work harvesting from thermal resources can be achieved by 
varying the particle statistics. We compare our results with 
previous investigations and discuss the effectiveness of our approach in Sec.~\ref{sec:discussion}.
We conclude in Sec.~\ref{sec:conclusion}.

%%%%%%%%%%%%%%%%%%%%%%%%%%%%%%%%%%%%%%%%%%%%%%%%%%%%%%%%%%%%%%%%%%%%%
\section{Working system: $q$-deformed oscillator}\label{sec:model}
\begin{figure}[t!]
	\centering
	\includegraphics[width=0.96\linewidth]{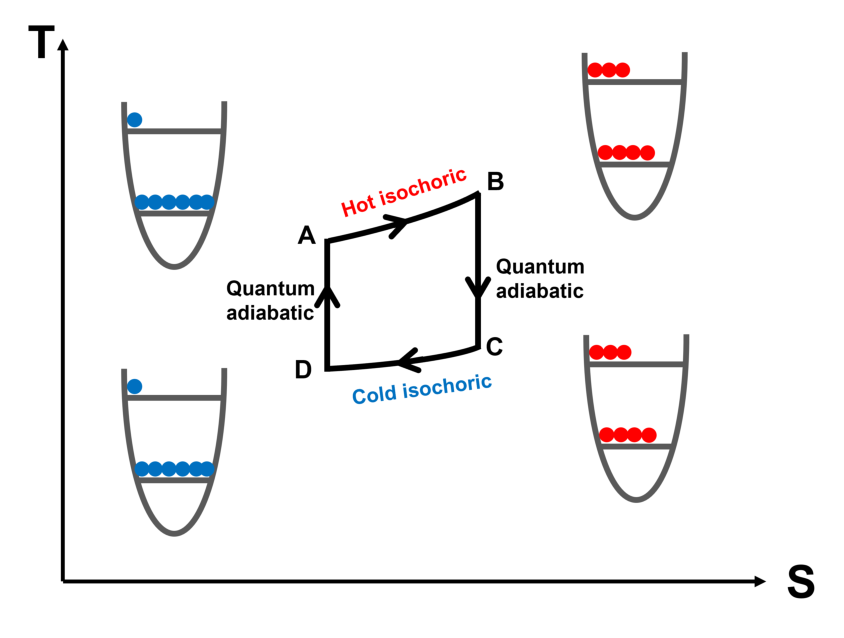}
	\caption{Entropy-temperature (S-T) diagram of the quantum Otto cycle. 
		In the isochoric processes, the system is in contact with hot and cold baths, while during the quantum adiabatic stages, quantum statistical character of the oscillator changes. 
	Populations, therefore the entropies remain the same during the quantum adiabatic stages, differentiating the quantum Otto cycle from its classical counterpart.}
	\label{fig:Q_otto_ss}
\end{figure}

%%%%%%%%%%%%%%%%%%%%%%%%%%%%%%%%%%%%%%%%%%%%%%%%%%%%%%%%%%%%%%%%%%%%%
We consider a q-deformed harmonic oscillator as the working substance of the Otto cycle as illustrated in Figure~\ref{fig:Q_otto_ss}.
The usual commutation relation of the 
Weyl-Heisenberg algebra of the quantum harmonic oscillator is
deformed in the case a $q$-deformed quantum oscillator according to~\cite{naseri2022,boumali2017statistical}
\begin{equation}
	\left[\hat a,\hat a^{\dagger}\right]_{q}=\hat a\hat 
	a^{\dagger}-q^{-1}\hat a^{\dagger}\hat a=q^{\hat N}\label{eq:deformedAlgebra}
\end{equation}
%where $\hat a$ and $\hat a^\dagger$ are the annihilation and creation operators, $q$ is the deformation parameter, and $\hat N=\hat a^\dagger \hat a $ is the number operator. 
where ${\hat N}$ is a number operator with eigenstates $|n\rangle$ such that ${\hat N} |n\rangle = n |n\rangle$, $a$ and $a^{\dagger}$ are lowering and raising operators in the spectrum of ${\hat N}$ such that $a^{\dagger} a |n\rangle = [n]|n\rangle$, and $q$ is the deformation parameter.

In terms of $\hat a$ and $\hat a^\dagger$,
the Hamiltonian of the $q$-deformed quantum oscillator with natural frequency $\omega$ is written as
\begin{equation}
	\hat H=\frac{\hbar\omega}{2}\left(\hat aa^{\dagger}+\hat a^{\dagger}\hat a\right).
	\label{eq:deformedOscHam}
\end{equation}
Using the $q$-number notation
\begin{equation}
	[n]=\frac{q^{n}-q^{-n}}{q-q^{-1}}\, ,\label{eq:qNumber}
\end{equation}
the eigenenergies of the Hamiltonian~(\ref{eq:deformedOscHam}) can be expressed in the form
\begin{equation}
	E_{n}=\frac{\hbar\omega}{2}\left(\left[n\right]+\left[n+1\right]\right).\label{eq:eigenenergies}
\end{equation}
These eigenenergies are associated with the deformed Fock number eigenstates
\begin{equation}
	\left|n\right\rangle =\frac{\left(a^{+}\right)^{n}}{\sqrt{\left[n\right]!}}\left|0\right\rangle ,\label{eq:eigenstate}
\end{equation}
where the $q$-factorial is defined to be $[n]!=[n][n-1]\cdot\cdot\cdot[1]$.

	With a real valued $r$, the deformation parameter can be considered to be a pure phase factor as $q = \text{exp}(i r)$, or to be a real number as $q = \text{exp}(r)$~\cite{bonatsos1999quantum}.
	 
\begin{figure}[b!]
	\centering
	\includegraphics[width=0.9\linewidth]{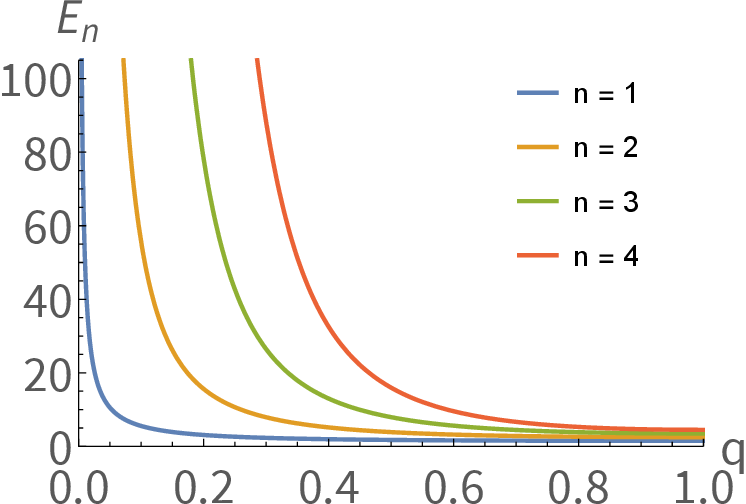}
	\caption{Energy eigenvalues as a function of the deformation parameter $q$ for energy eigen index $n=1,2,3,4$.}
	\label{fig:En}
\end{figure}

    Here we focus on the nonlinear characteristics of the quantum oscillator associated with $q$-deformation, and the deformation of bosonics system, hence we consider only real values with $0 < q < 1$~\cite{bonatsos1999quantum,boumali2017statistical}.
	The usual quantum harmonic oscillator relations can be recovered by substituting $q=1$ corresponding to the non-deformed case.
\begin{equation}
	E_{n}=\frac{\hbar\omega}{2}\frac{\sinh\left[ r \left(n+\frac{1}{2}\right)\right]}
	{\sinh\left(\frac{r}{2}\right)},
	\label{eq:simplifiedEigenenergy}
\end{equation}	% () changed to []
where $r=\ln{(q)}$ and $n = 0, 1, … < \infty $ is the energy eigen index.
$E_n$ is plotted in Figure~\ref{fig:En} as a function of $q$ for $n=1,2,3,4$.

Deformation parameter $q$ gives strong nonlinear character to the quantum oscillator, reflected in the anharmonicity of the energy spectrum. Without the deformation ($q=1$), quantum Otto cycle
can only be implemented by changing $\omega$ in the quantum adiabatic stages. When $\omega$ is increased,
the uniform energy gaps ($\hbar \omega$) between the eigenenergies increase, which can be compared to the
case of compressing the volume of the oscillator. In contrast, if $\omega$ is decreased, the energy gaps shrink as if the
volume of the oscillator is increasing. Accordingly, an effective piston like behavior can be translated to the quantum
oscillator by $\omega$ variation. For the case of a deformed quantum oscillator ($q \neq 1$), oscillator
has non-uniform energy gaps and their splitting can be further controlled by $q$. By keeping $\omega$ constant,
we can induce a piston-like behavior to a deformed oscillator by changing its quantum statistics per se.
Such a $q$-deformed oscillator uses $q$ as the quantum adiabatic control (expansion and compression) parameter
and it can exploit the anharmonic energy spectrum (for more expansion and compression effects) to enhance
the work harvesting and efficiency. Furthermore, alternatively we can still keep $q$ parameter constant and vary
$\omega$ as usual, but determine a critical quantum statistics (critical $q$) for which the work harvesting
and (or) efficiency would be maximum. 

\begin{figure}[t!]
	\centering
	\includegraphics[width=1\linewidth]{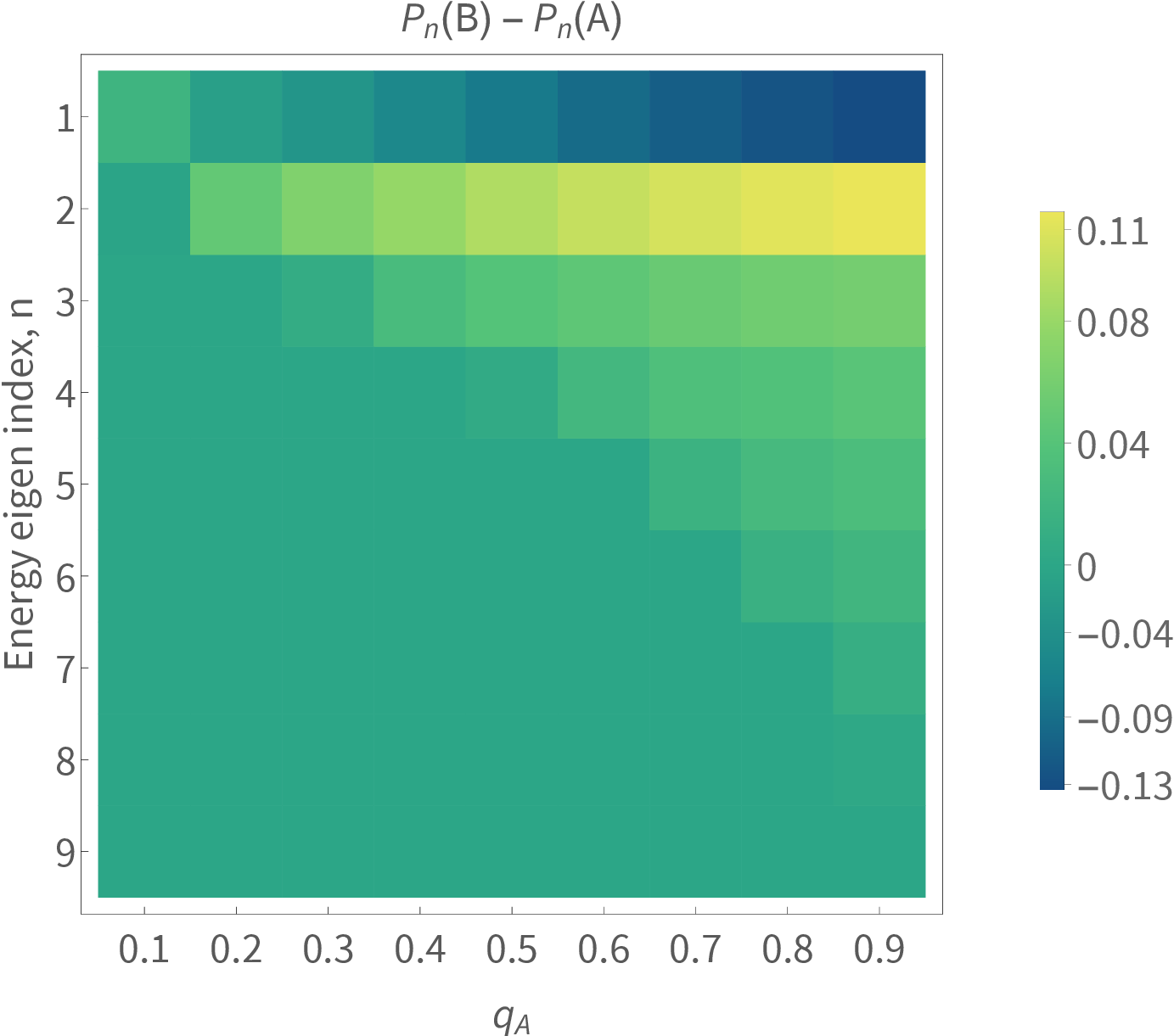}
	\caption{
		Population differences $P_n(B)-P_n(A)$ for $n$ and $q_A$ at $T_h = 0.5$ with $q_C = 1$ at $T_c = 0.1$.	}
	\label{fig:figPopDifs}
\end{figure}

%%%%%%%%%%%%%%%%%%%%%%%%%%%%%%%%%%%%%%%%%%%%%%%%%%%%%%%%%%%%%%%%%%%%
\section{Quantum Otto Cycle}\label{sec:quantumOttoCycle}
%%%%%%%%%%%%%%%%%%%%%%%%%%%%%%%%%%%%%%%%%%%%%%%%%%%%%%%%%%%%%%%%%%%%
The canonical partition function of the working system with Hamiltonian $\hat H$ in 
Eq.~(\ref{eq:deformedOscHam}) at temperature $T$ 
is given by
\begin{eqnarray}
	Z = \text{Tr}\left[\text{exp} (-\hat H/T)\right]=\sum_n \mathrm{e}^{-E_n/T},
\end{eqnarray}	% {} changed to []
where we take the Boltzmann constant $k_B$ as unity. Such a non-deformed structure of the partition 
function relies on the assumption of the non-deformed Gibbsian form of the thermal
equilibrium state $\rho \sim \text{exp} (-\beta\hat H)$ with $\beta $ = 1/T and 
the expectation value $\langle \hat A\rangle =\text{Tr}(\rho \hat A)$; and it is associated
with the assumption that Boltzmann-Gibbs form of the entropy function $S = \log W$ 
is preserved~\cite{tsallis88} where $W$ is the work output.

The quantum version of the classical Otto cycle~\cite{dittman2020heat} has been experimentally 
realized with quantum working substances~\cite{peterson2019, bouton2021}. 
As illustrated in Figure~\ref{fig:Q_otto_ss}, it consists of two quantum adiabatic and two isochoric heating/cooling stages. 
We can describe the cycle
in energy-population space for a quantum number $n$. At point $A$, the system is given 
with energy levels 
$E_n(A)$ and their populations $P_n(A)$. Under isochoric heating, the system 
is transformed ($A \rightarrow B$) 
to a thermal state at point $B$ such that $E_n(B)=E_n(A)$ and population is changed to $P_n(B)$. 
Subsequently,
the bath is removed and system is quantum adiabatically transformed ($B \rightarrow C$) obeying the
conditions $P_n(C)=P_n(B)$ while the energy eigenvalues are adiabatically changed to $E_n(C)$. In the third stage, the 
system is
isochorically cooled ($C \rightarrow D$) to point $D$ such that $E_n(D)=E_n(C)$ while the population 
is modified to $P_n(D)$.
Finally, the system is reset to its starting point $A$, ($D \rightarrow A$), by closing the 
cycle with another
quantum adiabatic transformation under the condition $P_n(A)=P_n(D)$.
\begin{figure}[t!]
	\centering
	\includegraphics[width=1\linewidth]{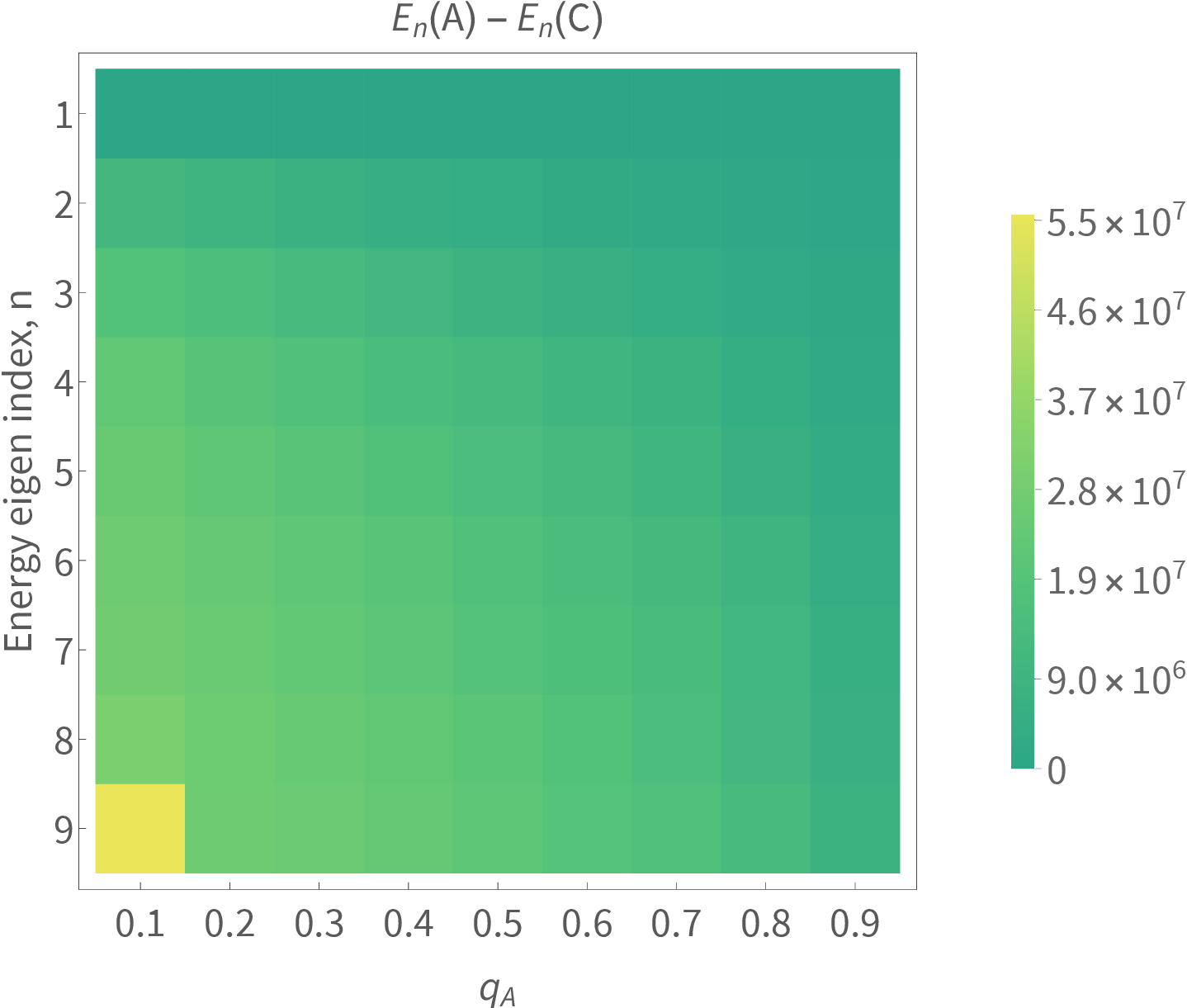}
	\caption{
		Net variation of energy eigenvalue $E_n(A)-E_n(C)$ during the adiabatic transformations for $n$ and $q_A$ at $T_h = 0.5$ with $q_C = 1$ at $T_c = 0.1$.	}
	\label{fig:figGapDifs}
\end{figure}

In classical Otto cycles, the control parameter in the isentropic steps is typically the 
physical volume
of the system. Here, we consider two types of quantum Otto cycles: (i) We keep the frequency 
of the quantum oscillator $\omega$ constant and vary the deformation parameter $q$ during 
the quantum adiabatic steps. (ii) We keep $q$ constant
and change $\omega$ in the quantum adiabatic stages. During these steps, the system is 
uncoupled from the
thermal baths and the system is quantum adiabatically transformed such that occupation 
probabilities $P_n$
of the eigenenergies $E_n$ do not change.
%(ii) We keep $q$ constant and change $\omega$ in the quantum adiabatic stages. 
During these steps, the system is 
uncoupled from the
thermal baths and the system is quantum adiabatically transformed such that occupation 
probabilities $P_n$
of the eigenenergies $E_n$ do not change. 

For a classical
heat engine, it is sufficient to make the transformation faster than the rate of heat exchange, 
instead of physically uncoupling the system from the environment, to ensure adiabatic condition. In the quantum case, transformation needs to be slower 
than characteristic time scale
for the transitions between the energy levels to ensure $P_n$ remains the same. 
Quantum case is also
an isentropic process, while the classical case does not necessarily satisfy the constant 
population condition. 
During the other two steps, isochoric heating and cooling, thermal baths are coupled 
to the oscillator system
while the parameters of the Hamiltonian and the quantum statistical parameter $q$ 
are kept constant. %\ \\ \ \\

\begin{figure}[t!]
	\centering
	\includegraphics[width=1\linewidth]{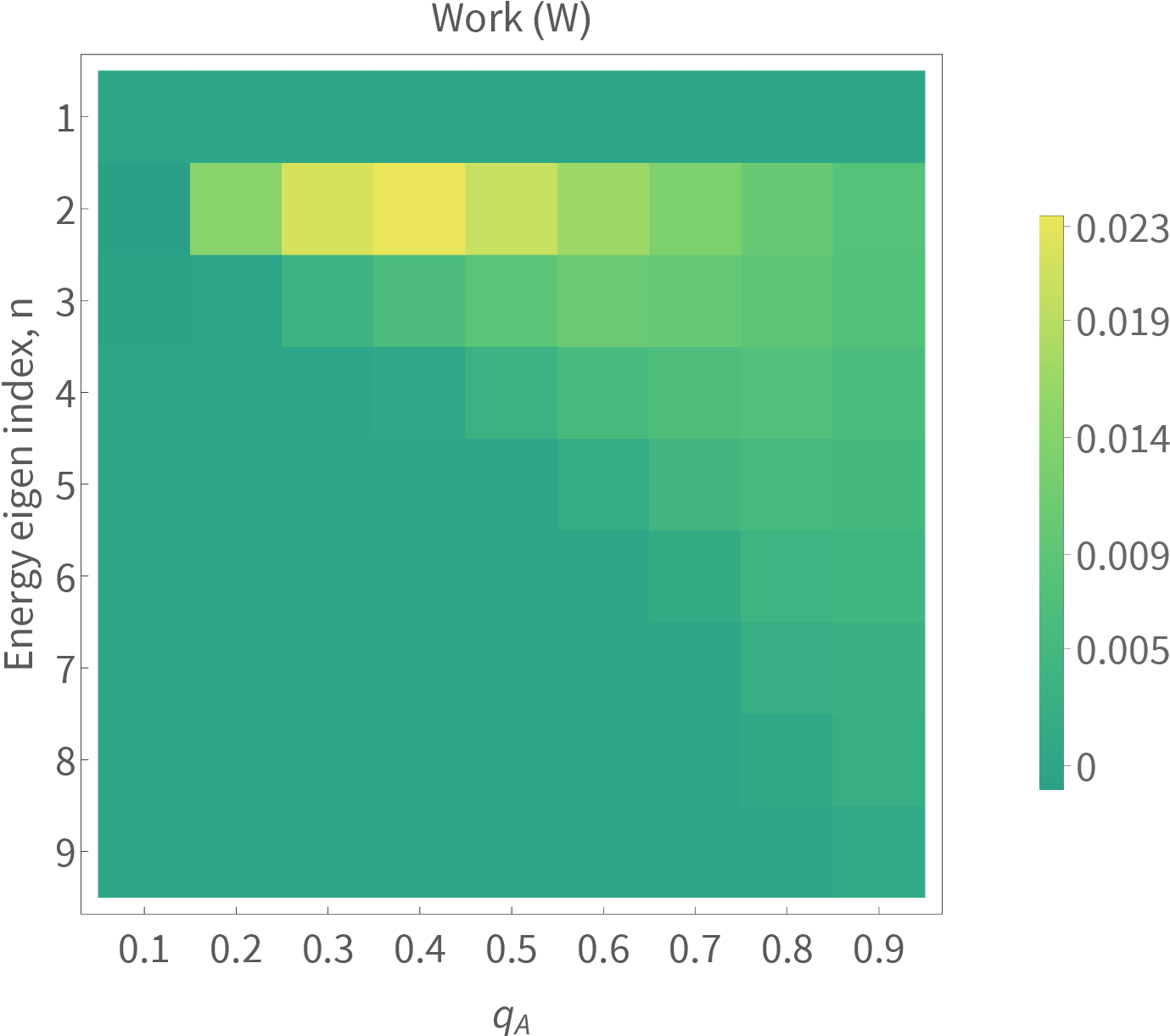}
	\caption{
		Extractable work for $n$ and $q_A$ at $T_h = 0.5$ with $q_C = 1$ at $T_c = 0.1$.	}
	\label{fig:figWorks}
\end{figure}

Using the energy eigenvalues $E_n$ in Eq.~(\ref{eq:eigenenergies}) and 
associated populations 
\begin{equation}
P_n = \exp(-E_n/T)/Z,	% {} removed
\end{equation}
we evaluate the work output ($W$) of the quantum Otto cycle according to the formula~\cite{quan2007}
\begin{equation}\label{eq:W}
W=\sum_n \left[E_n(A)-E_n(C)\right] \left[(P_n(B)-P_n(A)\right].
\end{equation}
$P_n(A)$ is the population of $n$-th energy level at the beginning of the 
isochoric heating. We can
use the quantum adiabatic condition $P_n(A)=P_n(D)$ to evaluate it using  
by the thermal distribution as the system is in thermal equilibrium at point 
$D$. $P_n(B)$ is calculated from 
a thermal distribution as well, as it is the level population at the end of 
isochoric heating. The
factor $E_n(A)-{\color{black}E_n(C)}$ is the net variation of the energy eigenvalue 
during the quantum 
adiabatic transformations.

Thermal efficiency of the engine is defined by
\begin{equation}
\eta=\frac{W}{Q_\text{in}}.
\end{equation}

\noindent 
Here, the injected heat into the system is given by
\begin{equation}
Q_\text{in}=\sum_n E_n(A)\left[P_n(B)-P_n(A)\right],
\end{equation}
where again we can use $P_n(A)=P_n(D)$ to evaluate it using thermal population 
distribution at point $D$.

%%%%%%%%%%%%%%%%%%%%%%%%%%%%%%%%%%%%%%%%%%%%%%%%%%%%%%%%%%%%%%%%%%%
\section{RESULTS}\label{sec:results}
%%%%%%%%%%%%%%%%%%%%%%%%%%%%%%%%%%%%%%%%%%%%%%%%%%%%%%%%%%%%%%%%%%%

%%%%%%%%%%%%%%%%%%%%%%%%%%%%%%%%%%%%%%%%%%%%%%%%%%%%%%%%%%%%%%%%%%%
\subsection{Work harvesting by statistical mutation}
%%%%%%%%%%%%%%%%%%%%%%%%%%%%%%%%%%%%%%%%%%%%%%%%%%%%%%%%%%%%%%%%%%%

\begin{figure}[t!]
	\centering
	\includegraphics[width=1\linewidth]{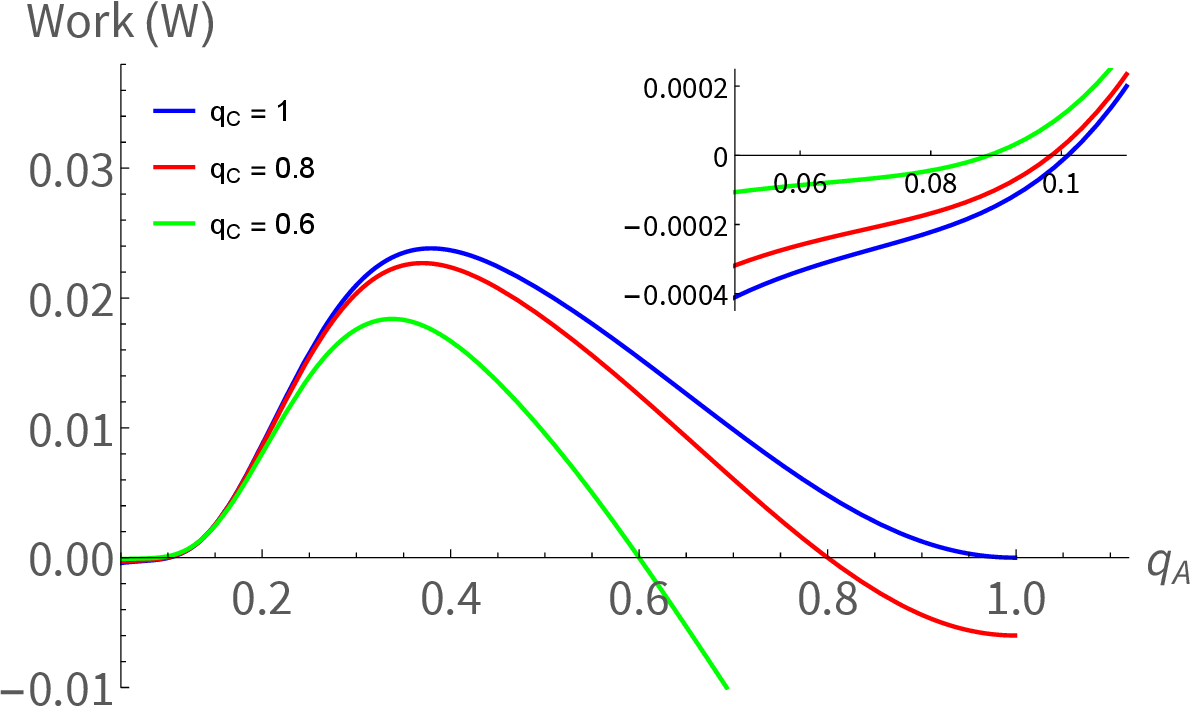}
	\caption{Work ($W$) by only changing the quantum statistics of the working substance ($q_A$) for  the hot isochore ($T_h=0.5$) for three choices of $q_C$ for the cold isochore ($T_c=0.1$) with fix oscillator frequency $\omega = 1$. 
	Global maxima for $W$ are at $q_A=0.379$, $q_A=0.369$ and $q_A=0.338$, respectively for the first three choices in descending order of $q_C$.
	{\color{black}
	  The inset shows the domain that positive work condition is not satisfied, i.e., $q_A < 0.101$, $q_A < 0.984$, and $q_A < 0.887$, respectively, for $q_C = 1$, $q_C = 0.8$, and $q_C = 0.6$.}	}
	\label{fig:qWork}
\end{figure}

Let us start changing the deformation parameter in the quantum adiabatic stages of the cycle.
For a q-deformed quantum oscillator, a decrease in $q$ yields an exponential increase in the energy gaps at higher energy levels. 
Thus, we take a smaller $q$ value ($q_A$) for hot isochore ($T_h$) and a higher $q$ value ($q_C$) for cold isochore ($T_c$). During the isentropic stages, $q$ varies between the values used in hot and cold isochores.
Hamiltonian parameter remain fixed at $\omega=1$, and only the $q$-deformation parameter of 
the working substance is varied in the cycle. 

Population differences $P_n(B)-P_n(A)$, the factor $E_n(A)-E_n(C)$, and their multiplication in Eq.~(\ref{eq:W}) for $n$-th energy level and $q_A$ at $T_h = 0.5$ are presented in Figures~\ref{fig:figPopDifs},~\ref{fig:figGapDifs} and~\ref{fig:figWorks}, respectively, for $q_C = 1$ at $T_c = 0.1$.
Here, $q_C = 1$ corresponds to non-deformed case showing that positive work can also be extracted by changing the statistics of the substance only for the hot isochore.
	Efficiency is meaningful and well-defined only when this positive work condition is satisfied.
	Analytical expression for the positive work condition could be possible if the energy gaps would change uniformly with $q$~\cite{quan2007}, which is not the case for our deformed oscillator. 
	Accordingly, we resort the numerical calculations of work for a range of temperature and $q$ to determine the positive work domains.

\begin{figure}[t!]
	\centering
	\includegraphics[width=1\linewidth]{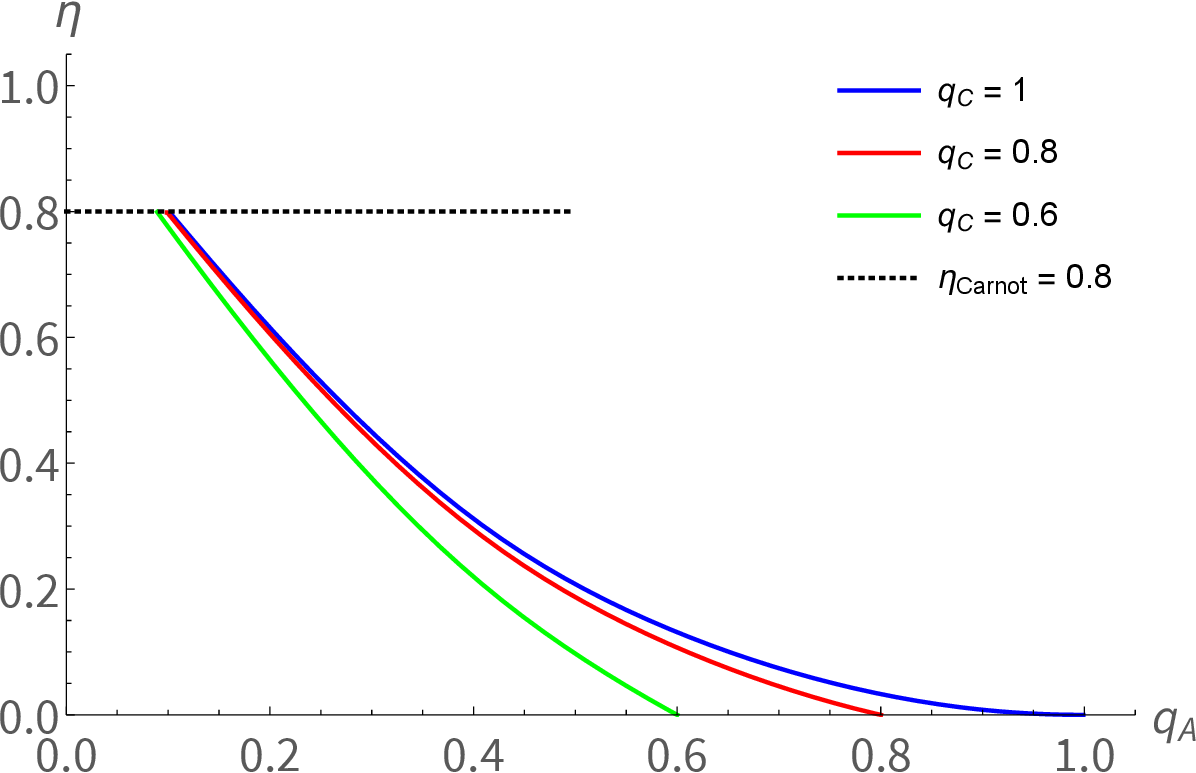}
	\caption{			Efficiency of quantum Otto cycle as a function of $q_A$ (at $T_c = 0.1$) with three choices of $q_C$ (at $T_h = 0.5$).
			The functions are plotted not for the entire range of $q_A$ but only in the positive work domain as presented in the inset of Figure~\ref{fig:qWork}. As $q_A$ approaches its minimum within the positive work domain, the quantum Otto engine's efficiency $\eta$ approaches the Carnot limit $\eta_{\text{Carnot}} = 1 - T_c / T_h = 0.8$.
		}
	\label{fig:qEff}
\end{figure}

Summing the work values over $n$ in Figure~\ref{fig:figWorks}, the extractable work (W) with respect to $q_A$ is presented in Figure~\ref{fig:qWork} at $T_h = 0.5$ for three choices of $q_C$ at $T_c = 0.1$. 
{\color{black}
	Figure~\ref{fig:qEff} presents the efficiency as a function of $q_A$ with the same choices of $q_C$ and bath temperatures. 
	Quantum Otto engine's efficiency approaches the Carnot limit only by $q$-deformation as $q_A$ gets smaller within the positive work domain which is numerically found and shown in the inset of Figure~\ref{fig:qWork}. 
	Note that it is straightforward to check numerically for any choice of $q$ and bath temperatures that in the positive work condition, the Carnot limit is satisfied.
	For example, setting the hot isochore $T_h = 1$ with the same choices as in Figure~\ref{fig:qWork}, the positive work condition is satisfied until $q_A \approx 0.05$ at which the quantum Otto efficiency approaches the Carnot limit $\eta_{\text{Carnot}} = 0.9$.
}

\begin{figure}[b!]
	\centering
	\includegraphics[width=0.96\linewidth]{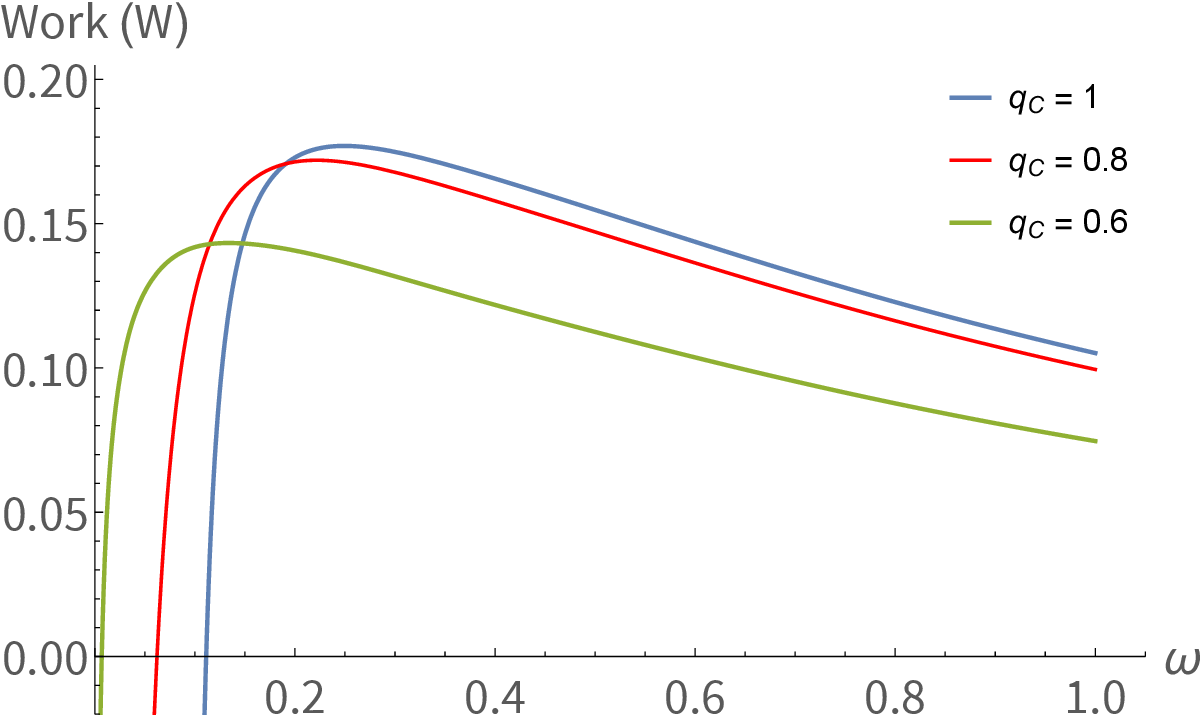}
	\caption{Extractable work as a function of $\omega$ with $q_A=0.4$ at $T_h=1$ and $q_C$ at $T_c=0.1$. For small $\omega$, an optimal choice of $q_C$ yields maximum extractable work.}
	\label{fig:qWorkOmegas}
\end{figure}

%%%%%%%%%%%%%%%%%%%%%%%%%%%%%%%%%%%%%%%%%%%%%%%%%%%%%%%%%%%%%%%%%%%
\subsection{Critical quantum statistics for optimal work harvesting}
%%%%%%%%%%%%%%%%%%%%%%%%%%%%%%%%%%%%%%%%%%%%%%%%%%%%%%%%%%%%%%%%%%%

For a given temperature difference between hot and cold reservoirs as a classical resource,
we can point out a critical quantum statistics of a deformed quantum oscillator
to maximize the harvested work or efficiency. Ref.~\cite{deffner2020} has shown that a 
bosonic system has higher engine performance than a fermionic system 
and stated that the performance difference occurs due to the difference in internal energies 
arising from the Pauli exclusion principle. 
In the same spirit, we find out that one can 
optimize the engine performance for a given thermal resource in terms of the quantum 
statistical character of the working substance. 
As shown in Figure~\ref{fig:qWork}, $q_A \approx 0.4$ is the optimal value for the working substance for the hot isochore.

We now keep $q_A = 0.4$ for various $q_C$ and induce the cycle for different $\omega$. As shown in Figure~\ref{fig:qWorkOmegas}, for $\omega<0.2$, there is an optimal $q_C$ that achieves maximum extractable work.
\begin{figure}[b!]
	\centering
	\includegraphics[width=0.96\linewidth]{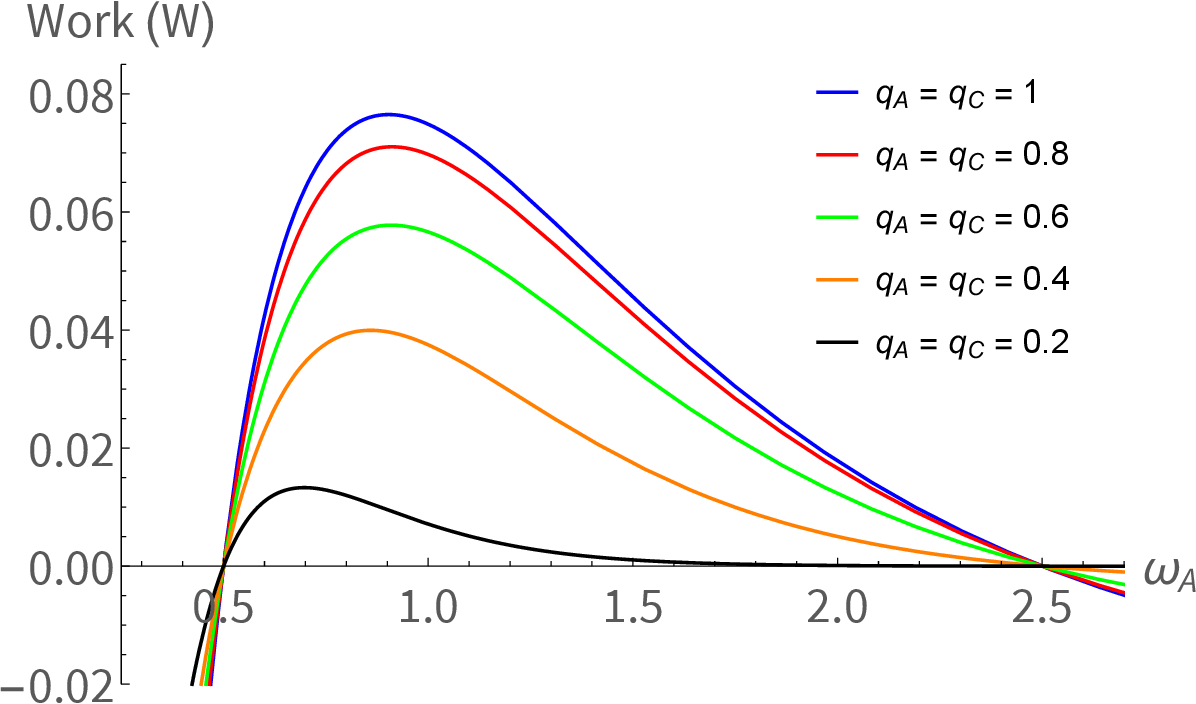}
	\caption{Extractable work as a function of $\omega_A$ with $\omega_C = 0.5$ with $T_c=0.1$ and $T_h=0.5$ with the substance deformed at same level in adiabatic steps ($q_A=q_C$).}
	\label{fig:qWorkOmegaA}
\end{figure}
Lastly, for the second type of quantum Otto engine, we keep $q$ constant (i.e., $q_A = q_C$) and change $\omega$ in the quantum adiabatic steps, i.e., we choose different $\omega_A$ and $\omega_C$.
As can be seen in Figure~\ref{fig:qWorkOmegaA}, greater $q$ implies greater extractable work and maximum in the nondeformed case ($q_A = q_C = 1$). 
This result shows that the advantage of $q$-deformation with no classical analogue is due not to a simply deformed working substance, but rather to utilizing the working substance with different deformation levels at the adiabatic stages even with the same $\omega$.

%%%%%%%%%%%%%%%%%%%%%%%%%%%%%%%%%%%%%%%%%%%%%%%%%%%%%%%%%%%%%%%%%%%
\section{Discussion}\label{sec:discussion}
%%%%%%%%%%%%%%%%%%%%%%%%%%%%%%%%%%%%%%%%%%%%%%%%%%%%%%%%%%%%%%%%%%%

Let us consider an anharmonic Hamiltonian
\begin{equation}
	\hat H_a = \mu_1 a^{\dag} a + \mu_2 a^{\dag} a a^{\dag} a
\end{equation}
\noindent
where nonlinearity can be represented by deformation as
\begin{equation}
	a_q = a + {\epsilon \over 4} a a^{\dag} a
  \label{eq:nonlinearity}
\end{equation}
\noindent with $\epsilon = 1 - q \ll 1$.
Then, as shown in~\cite{birol2009phase}, Kerr medium or s-wave scattering in atomic BEC can be simulated through nonlinear interactions of the deformed oscillator.

It is found in accordance with previous works~\cite{ivanchenko2015quantum,thomas2011coupled,ccakmak2017special,ferdi2014,boubakour2023interaction} that the extractable work and efficiency are not monotonic with respect to the interaction parameter, but rather exhibit optimum values as presented in Figure~\ref{fig:qWork}.

	We remark that we do not include explicitly the work reservoir into the engine model as it is a common treatment in quantum heat engines.
	It was shown in~\cite{li2018efficient} that in a thermodynamic cycle using particles with nonlinear interactions as the working substance, adjusting the Feshbach resonances~\cite{feshbach1958unified} can tune the nonlinear interaction strength to produce work by modifying the volume of the gas. 
	Similarly in our model, it should be understood that the work of the engine will be done against the magnetic field used to modify Feshbach resonances to change the $q$-parameter associated with the nonlinearity of the working substance.

\begin{figure}[t!]
	\centering
	\includegraphics[width=0.96\linewidth]{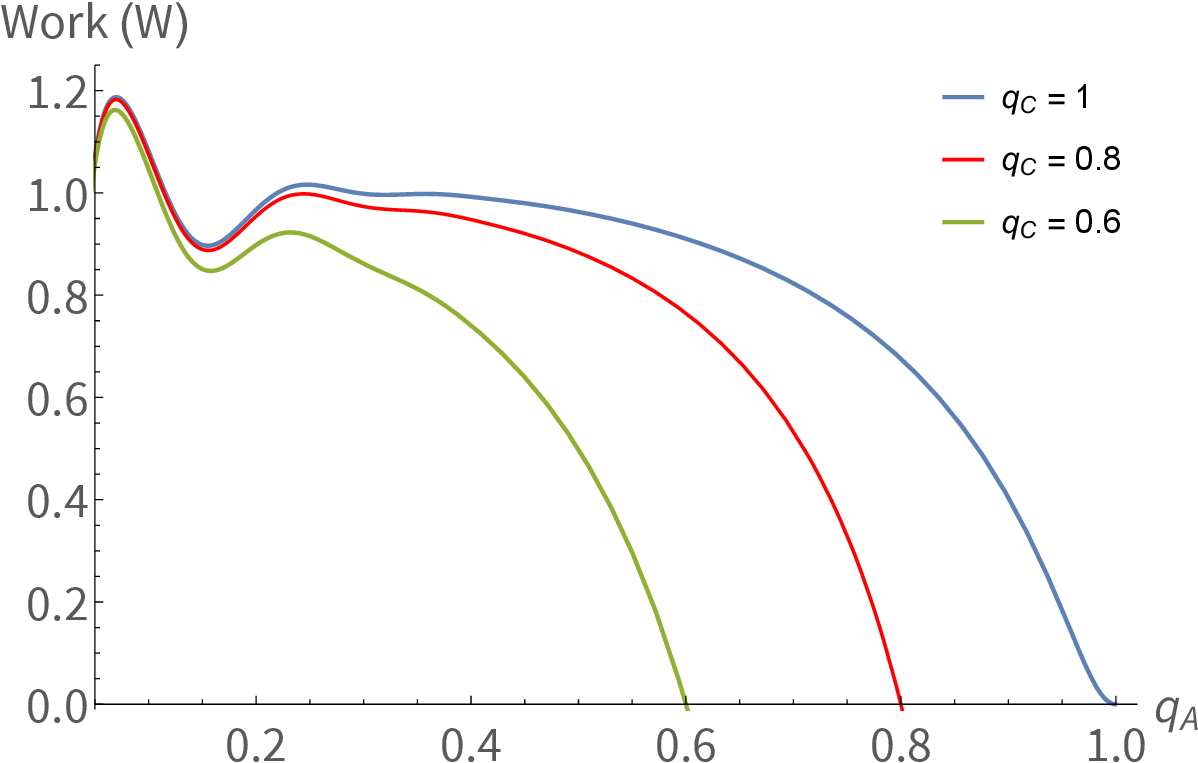}
	\caption{Work with respect to $q_A$ with a greater temperature difference between hot isochore ($T_h=5$) and  cold isochore ($T_c=0.1$).}
	\label{fig:qWorkTHot5}
\end{figure}

\begin{figure}[t!]
	\centering
	\includegraphics[width=0.96\linewidth]{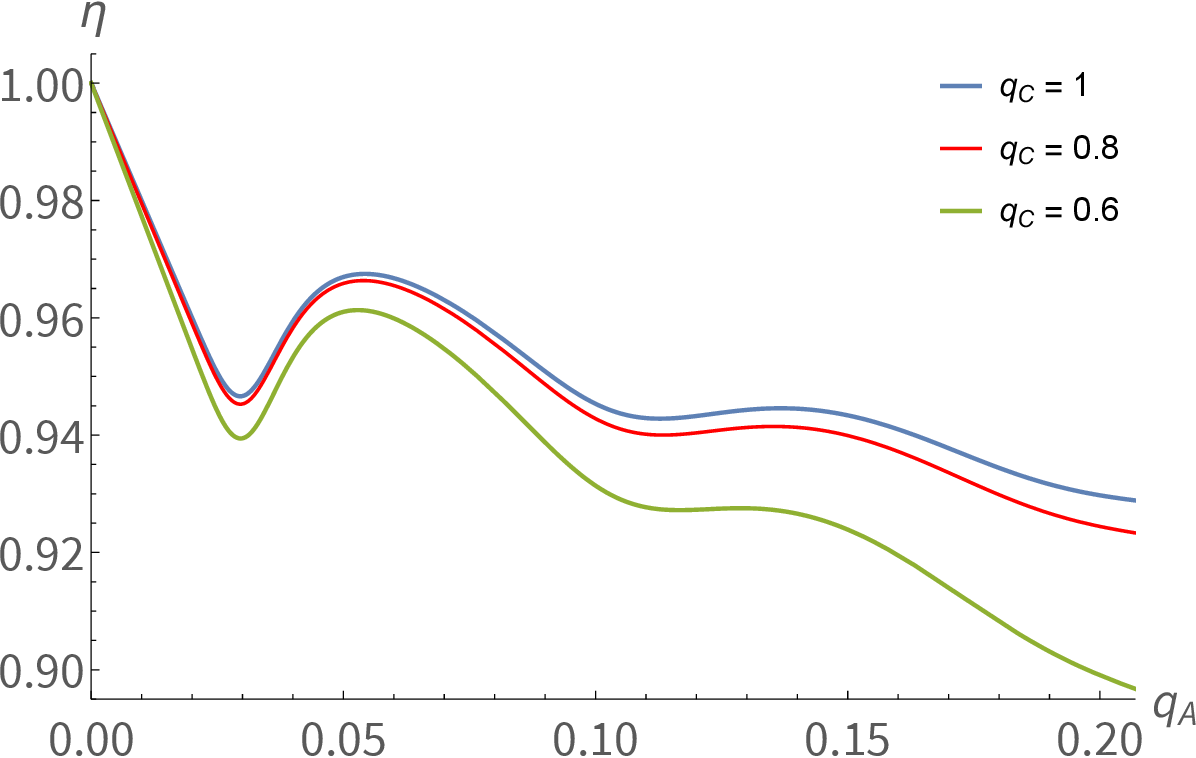}
	\caption{Efficiency of quantum Otto cycle as a function of $q_A$ at $T_c = 0.1$ and three choices of $q_C$ at $T_h = 100$  with $\omega = 1$.}
	\label{fig:qEffTHot100}
\end{figure}

The effect of non-linearity due to $q$-deformation can be observed more clearly with greater temperature difference between the cold and the hot bath as presented in Figures~\ref{fig:qWorkTHot5} and~\ref{fig:qEffTHot100} for work and efficiency, respectively.
	Note that due to weak nonlinearity achievable in experiments, we considered a small $\epsilon$ in Eq.~(\ref{eq:nonlinearity}). 
	However, considering that strong nonlinearities and other physical phenomena could also be associated with $q$-deformation, we present our results for the entire range of $q$-deformation parameter in the positive work domain satisfying the Carnot limit.

	Algebras corresponding to both $q$-deformed fermions and $q$-deformed bosons, as well as their transformations and unification are widely studied~\cite{bonatsos1992generalized,bonatsos1999quantum,narayana2001transformations,sargolzaeipor2019q,lavagno2002}.
	Considering potential realizations in atomic BEC systems, also in accordance with the findings of Ref.~\cite{deffner2020} that bosonic working system achieves a higher performance than fermionic systems in the non-deformed setting ($q=1$), in this work we focused on $q$-deformed bosonic system as the working substance. 
	Nevertheless, it would be interesting as future research to study different performance characteristics of complex deformations, and $q$-deformed fermionic systems as well, especially following the unified representation by Lavagno and Swamy~\cite{lavagno2002}. 

In another work, the effect of $q$-deformation was studied to show a relationship 
between the efficiency of QHE and the non-Markovianity in the engine cycle~\cite{naseri2022}.
It states that the $q$-parameter, which causes non-equilibrium dynamics, helps to build a 
relationship between theoretical and experimental results. 
Here, we did not examine finite time engine cycles; the $q$ parameter, in our case, plays 
a more active and direct role as the engine cycle's control parameter. 

%%%%%%%%%%%%%%%%%%%%%%%%%%%%%%%%%%%%%%%%%%%%%%%%%%%%%%%%%%%%%%%%%%%
\section{CONCLUSION}\label{sec:conclusion}
%%%%%%%%%%%%%%%%%%%%%%%%%%%%%%%%%%%%%%%%%%%%%%%%%%%%%%%%%%%%%%%%%%%
Classical heat engines harvest work from a thermal resource by converting a heat flow 
between a hot and cold bath to an ordered work. To realize this task, parameters of 
their classical working system, for example, the volume of working gas, are varied in an engine 
cycle. Quantum heat engines use a quantum working material, and in addition to external degrees 
of freedom, Hamiltonian parameters and internal degrees of freedom can also be utilized for 
work extraction. Profound quantum effects, particularly improving engine performance over 
its classical counterpart, such as via quantum correlations~\cite{tuncer2019work}, are possible with quantum heat engines. 

Here, we show that the quantum statistical character of the working substance can be 
used as another control parameter of a quantum heat engine. Specifically, we consider a 
quantum oscillator with a fixed frequency $\omega$ but deformed quantum statistics characterized 
by the $q$-deformation parameter and verify that such a $q$-deformed oscillator can harvest 
work from a thermal resource by variation of the $q$ parameter. Alternatively,
we can optimize the harvested work or efficiency for a given thermal resource by choosing a 
critical quantum
statistics of the deformed oscillator. 

While we consider the Otto cycle as a paradigmatic model, we expect that our fundamental conclusion 
holds for other engine cycles as well. Variation of particle statistics can be experimentally 
challenging relative to the traditional way of variation of Hamiltonian parameters. 
However, $q$-deformation can be envisioned and mapped to nonlinear terms in Hamiltonians 
and effective engineered deformed oscillator models ranging from semiconductor 
cavity QED~\cite{liu2001} to atomic Bose-Einstein condensates~\cite{gardiner1997} 
could be explored for physical embodiment of the statistical mutation route of work extraction.

%%%%%%%%%%%%%%%%%%%%%%%%%%%%%%%%%%%%%%%%%%%%%%%%%%%%%%%%%%%%%%%%%%%
\acknowledgements
%\section*{acknowledgements}\label{sec:ack}
%%%%%%%%%%%%%%%%%%%%%%%%%%%%%%%%%%%%%%%%%%%%%%%%%%%%%%%%%%%%%%%%%%%
This study was funded by Istanbul Technical University BAP-41181. F.O. acknowledges the Personal Research Fund of Tokyo International University.

%\vspace{-1cm}

%%%%%%%%%%%%%%%%%%%%%%%%%%%%%%%%%%%%%%%%%%%%%%%%%%%%%%%%%%%%%%%%%%%
%\bibliographystyle{apsrev4-2}
\bibliography{qDefOtto}
%%%%%%%%%%%%%%%%%%%%%%%%%%%%%%%%%%%%%%%%%%%%%%%%%%%%%%%%%%%%%%%%%%%

\end{document}